\documentclass[12pt]{aastex}
\usepackage{emulateapj5}
\submitted{ApJ}
\usepackage{psfig}
\newcommand{\mamo}[1]{\mbox{$#1$}}
\newcommand{\unit}[1]{\ifmmode \:\mbox{\rm #1}\else \mbox{#1}\fi}
\newcommand{\sbr}[1]{_{\rm #1}}

\newcommand{\mone}{\mamo{^{-1}}}

\newcommand{\ten}[1]{\mamo{\times 10^{#1}}}

\newcommand{\kms}{\unit{km~s\mone}}

\newcommand{\mpc}{\unit{Mpc}}

\newcommand{\hmpcinvcub}{\mamo{h_{75}^{3} \mpc^{-3}}}
\newcommand{\ls}{L\sbr{s}}
\newcommand{\lx}{L\sbr{X}}
\newcommand{\lgal}{L\sbr{gal}}

\def\lesssim{\mathrel{\hbox{\rlap{\hbox{\lower4pt\hbox{$\sim$}}}\hbox{$<$}}}}
\def\gtrsim{\mathrel{\hbox{\rlap{\hbox{\lower4pt\hbox{$\sim$}}}\hbox{$>$}}}}

\slugcomment{\apj,  submitted}

\begin{document}

\title{THE MASS--TO--LIGHT FUNCTION OF  VIRIALIZED SYSTEMS  AND \\
THE RELATIONSHIP BETWEEN THEIR OPTICAL AND X-RAY PROPERTIES}

\author{Christian MARINONI}
\affil{Department of Astronomy, University of California at Berkeley, Berkeley, CA 94720-3411, US}
\email{marinoni@astro.berkeley.edu}

\author{Michael J. HUDSON}
\affil{Department of Physics, University of Waterloo, Waterloo, Canada}
\email{mjhudson@astro.uwaterloo.ca}

\begin{abstract}

We compare the B-band luminosity function of virialized halos with the
mass function predicted by the Press-Schechter theory in cold dark
matter cosmogonies. We find that all cosmological models fail to match
our results if a constant mass--to--light ratio is assumed.  In order
for these models to match the faint end of the luminosity function, a
mass--to--light ratio decreasing with luminosity as $L^{-0.5\pm 0.06}$
is required.  For a $\Lambda$CDM model, the mass--to--light function
has a minimum of $\sim 100 h^{-1}_{75}$ in solar units in the $B$-band,
corresponding to $\sim 25\%$ of the baryons in the form of stars, and
this minimum 
occurs close to the luminosity of an $L^*$ galaxy.  At
the high-mass end, the $\Lambda$CDM model requires a mass--to--light
ratio increasing with luminosity as $L^{+0.5 \pm 0.26}$.  This scaling
behavior of the mass--to--light ratio appears to be in qualitative
agreement with the predictions of semi-analytical models of galaxy
formation.  In contrast, for the $\tau$CDM model, a constant
mass--to--light ratio suffices to match the high-mass end.

We also derive the halo occupation number, i.e. the number of galaxies
brighter than $\lgal^*$ hosted in a virialized system. 
We 
find that
the halo occupation number
scales non-linearly with the total mass of the 
system,
$N\sbr{gal}(>\lgal^*) \propto m^{0.55\pm0.026}$ 
for the $\Lambda$CDM model.

We find a break in the power-law slope of the X-ray-to-optical
luminosity relation, 
independent of the cosmological model. This break occurs 
at a scale corresponding to poor groups.  In the $\Lambda$CDM model,   
the poor-group mass is also the scale at which the mass-to-light ratio 
of virialized systems begins to increase.                              
This 
correspondence 
suggests a physical link between 
star 
formation and the X-ray
properties of halos, possibly due to preheating by supernovae or to
efficient cooling of low-entropy gas into galaxies.
\end{abstract}


\section{INTRODUCTION}

One of the major problems of cosmology is to determine the connection
between mass and light in the universe over different scales. 
The abundance by mass of virialized dark matter halos is a fundamental
prediction of cosmological models (Press \& Schechter 1974).
Observationally, halos must be identified with virialized galaxy
``systems'', ranging in size from single dwarf galaxies to rich
clusters of galaxies.

The masses, and hence the mass multiplicity functions, derived from
cataloged systems are not robust, particularly for small systems such
as
poor
groups which contain only a few galaxies.  In contrast, halo light is
a fundamental observable obeying a robustly determined distribution
function, the luminosity function of virialized systems (VSLF).

In a previous paper (Marinoni et al.\ 2001b, Paper V), we measured the
VSLF in the nearby Universe using an objective catalog of virialized
systems.  In this paper we show how the VSLF can be used to
investigate scaling relations between mass and light.

The direct approach to obtaining scaling relations between, for
example, mass and light is to regress one parameter on the other. 
This 
direct approach has several weaknesses, however. As noted above, 
for low mass systems 
the measured masses can have large errors.
Second, the catalogs may be incomplete in either mass or
light. Furthermore, it is usually necessary to patch together results
from several independent catalogs, each of which probes different mass
scales (e.g., galaxies, groups and clusters).

In this paper we take a different approach. We derive the mass
functions from theory and compare these with the observed 
VSLF,
solving for the unknown mass--to--light ratio ($\Upsilon$) as a
function of the total mass or luminosity of the halo. The price paid
in this approach is that the 
mass function depends 
on the assumed cosmological model.  The advantage over the
conventional approach is that we can simultaneously and seamlessly
explore a wide dynamic range in mass and luminosity,
free of systematics. 

The $\Upsilon$ function, which measures the efficiency with which the
universe transforms matter into light, can thus be used not only as a
traditional estimator of the density parameter $\Omega\sbr{m}$ but
also as a diagnostic tool for different models of galaxy formation.  

A closely related statistic, useful in modeling galaxy formation and
clustering, is the halo occupation number, i.e. the number of galaxies
above some luminosity threshold (chosen to be the absolute magnitude
of an $L^*$ galaxy) that populate a virialized halo as a function of
its total mass or luminosity.

Within the same framework, we can also investigate the emission
properties of halos in different wavelength domains of the
electromagnetic spectrum such as in the optical and X-ray bands. The
sample of 
systems 
for which both the optical and X-ray luminosities
are  well measured 
is still limited, and does not allow a reliable determination of the
scaling of the X-ray to optical light ratio. However, by comparing
statistically their luminosity distributions, we can constrain the
functional behavior of this ratio.

This paper is the sixth in a series (Marinoni et al.\ 1998, Paper I;
Marinoni et al.\ 1999, Paper II; Giuricin et al.\ 2000, Paper III;
Giuricin et al.\ 2001, Paper IV; Marinoni et al.\ 2001b, Paper V) in
which we investigate the properties of the large-scale structures as
traced by the NOG sample. The outline of our paper is as follows: in
\S 2 we review and summarize the determination of the observed
luminosity function of virialized halos.  In \S 3 we compare our
results with the predictions of $\tau$CDM and $\Lambda$CDM models, as
well as the predictions from semi-analytic models.  In \S 4, we
measure the halo occupation number.  In \S 5 we examine the
relationship between the optical and X-ray properties in virialized
systems. In \S 6 we discuss in detail the resulting scaling relations
between the properties of groups and clusters of galaxies.  Results
are summarized in \S 7.

Throughout this paper, the Hubble constant is taken to be 75 $h_{75}$
km s$^{-1}$ Mpc$^{-1}$ and the recession velocities $cz\sbr{LG}$ are
evaluated in the Local Group rest frame.

\section{The Luminosity Function of Virialized Halos }

In paper V, we determined the luminosity function of virialized
systems using groups extracted from the NOG galaxy catalog.  The NOG
sample (Marinoni 2001a) is a statistically controlled, distance-limited
($cz\sbr{LG}\leq$6000 \kms) and magnitude-limited ($B \leq 14$)
complete sample of more than $7000$ optical galaxies.  The sample
covers 2/3 (8.27 sr) of the sky ($|b|>20^{\circ}$), a volume of $1.41
\times 10^{6}\hmpcinvcub$ and has a redshift completeness of 98\%.

From the NOG, three different ``virialized system'' samples were
determined using groups selected by three different objective
group-finding algorithms (Paper III).  In paper V, we showed that the VSLF
derived from these different samples are consistent and hence that the
VSLF is robust.

In order to recover accurately the absolute luminosity of a given
halo, we correct for both the integrated luminosity of galaxies below
the magnitude limit of the sample and for the completeness of
subsamples of the halo catalog.  Moreover, models to correct for the
large-scale motions (Paper I) have been applied in order to improve
the determination of the halo distances.  The key steps of the
reconstruction procedure are shown in Fig. \ref{fig1}, focusing on
several virialized systems in the Virgo region which are embedded in
lower-density filamentary large-scale structure.

In Paper V, we found that the $B$-band luminosity function of the
whole hierarchy of gravitationally bound systems, from single galaxies
to rich clusters, was insensitive to the choice of the group-finding
algorithms with which halos are selected, and was well described over
the absolute-magnitude range $-24.5 \leq M\sbr{s} + 5 \log h_{75} \leq
-18.5$ by a Schechter function (Schechter 1976) with
$\alpha\sbr{s}=-1.4\pm 0.03$, $M\sbr{s}^{*} - 5 \log h_{75} =-23.1 \pm
0.06$ and $\phi\sbr{s}^{*}=4.8 \times 10^{-4}\;\hmpcinvcub$ or by a
double power law: $\phi_{\rm pl}(\ls) \propto \ls^{-1.45 \pm 0.07}$
for $\ls< L\sbr{pl}$ and $\phi(\ls) \propto \ls^{-2.35 \pm 0.15}$ for
$\ls > L\sbr{pl}$ with $L\sbr{pl}=8.5 \times 10^{10} h_{75}^{-2}
L_{\odot}$, corresponding to $M\sbr{s}- 5 \log h_{75} = -21.85$.  The
characteristic luminosity of virialized systems, $L\sbr{pl}$, is $\sim
3$ times brighter than that ($L^{*}\sbr{gal}$) of the luminosity
function of NOG galaxies.

\section{The Mass--to--Light Ratio of Virialized Systems}

\subsection{Introduction}

In order to compare observational data with theoretical predictions,
the traditional approach has been to estimate the halo mass of groups
and clusters directly from the data, using, for example, velocity
dispersions of galaxy members (Carlberg et al.\ 1996; Girardi et al.\
1998), X-ray gas temperatures (David, Jones, \& Forman 1996; Lewis et
al.\ 1999), and gravitational lensing (Smail et al.\ 1997, Allen
1998).

The application of these methods is reasonably straightforward for
rich clusters of galaxies but is more problematic for poor groups, due
to their low X-ray surface brightness and the poor sampling of optical
dynamical mass estimators (see, e.g., Zabludoff \& Mulchaey 1998,
Mahdavi et al.\ 1999).  Consider, for example, a system composed of a
few elements selected from a magnitude-limited survey.  In this case
one will miss faint members that do not satisfy selection criteria.
The contribution of faint members to the total luminosity of the
system is negligible small, and can easily be corrected (see Paper V).
In contrast, these faint galaxies are important dynamical probes of
the gravitational potential well, and their omission can translate
into a serious bias
in the estimated mass, 
particularly if galaxies of different luminosities are clustered in
different ways (Park et al.  1994; Giuricin et al.\ 2001).  Moreover
one can construct a completeness function and than derive a
differential distribution function for the light emitted by virialized
systems, while the construction of a similar completeness function
involving dynamical estimators (velocity dispersion for example) is
very problematic (Borgani et al. 1997).  Thus, the luminosity function
of virialized systems is more robust on group scales than is the mass
function derived using projected velocity dispersions.

In this paper, following Cavaliere, Colafrancesco \& Scaramella (1991)
and Moore, Frenk, \& White (1993, MFW), we explore an alternative
approach to determining the mass--to--light ratio over a range of
scales including both poor and rich systems.  Specifically, we compare
the robust observationally-determined luminosity function of systems
to theoretically-determined mass functions.

\subsection{Theory}

The differential mass function of virialized systems $n(m)$ can be
calculated, for a given cosmological model, via the Press-Schechter
formalism (Press \& Schechter 1974, hereafter PS; Schaeffer \& Silk
1985; Bower 1991; Bond et al.\ 1991; Cavaliere, Colafrancesco, \&
Scaramella 1991; Lacey \& Cole 1993; Monaco 1998), or more accurately,
through N-body simulations (Efstathiou \& Rees 1988; Lacey \& Cole
1994; Governato et al.\ 1999; Jenkins et al.\ 2001).  The comparison
with N-body simulations reveals that the PS mass function provides a
reasonably accurate description of the abundance of virialized halos
on group and cluster scales although it tends to overpredict 

\vbox{%
\begin{center}
\leavevmode
\hbox{%
\epsfxsize=19.9cm
\epsffile{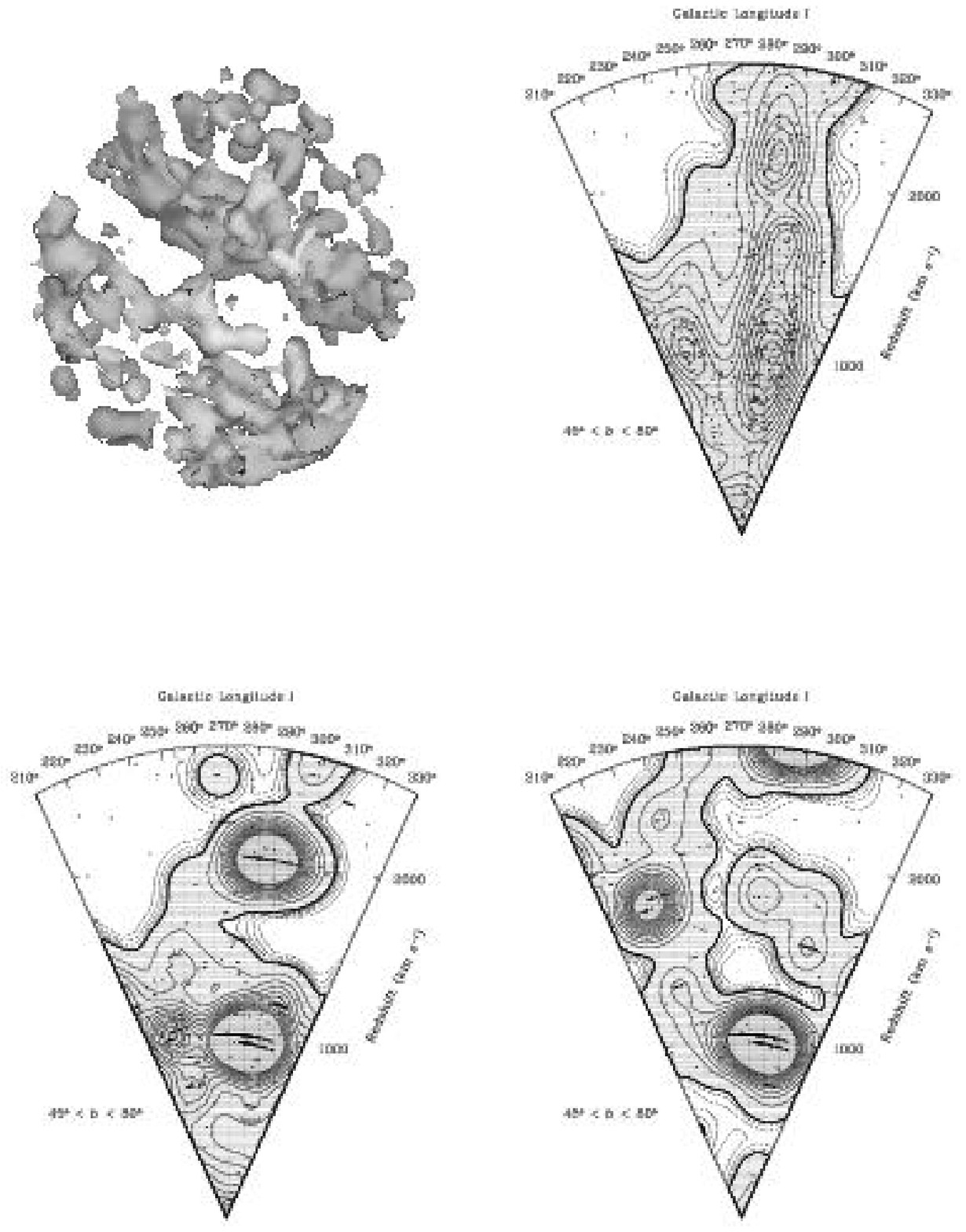}}
\begin{small}
\figcaption{%
{\em Upper left:} Density ($\delta$=1.5) distribution of NOG galaxies
smoothed using a Gaussian window function with smoothing length
$R\sbr{s}=200$ \kms.  {\em Upper right:} redshift distribution of NOG
galaxies in a cone diagram centered on the Virgo region (smoothed with
$R\sbr{s} = 100 \kms$).  Regions with $\delta>0$ are in grey scale
while contours are spaced by $\Delta \delta=2$ {\em Lower left:} the
same region after clustering reconstruction with the hierarchical
method. ({\em Lower right:} the same region after having applied the
peculiar velocity field model fitted using Mark III data.
\label{fig1}}
\end{small}
\end{center}}
\newpage

(albeit
by a small amount) the abundance of low-mass halos and underpredict
that of high-mass halos (e.g., Gross et al.\ 1998; Sheth \& Tormen
1998; Governato et al.\ 1999).  Given the uncertainties in the
luminosity function of systems to which we will be comparing the mass
function, the PS formula is of sufficient accuracy.

The PS analytical expression for the present-day comoving number
density of dark halos of mass $m$ in the interval $dm$ is

\begin{equation}
n(m)dm=\sqrt{\frac{2}{\pi}} 
\frac{\overline{\rho}\delta_c}{\sigma(m)^2 m} 
\exp 
\left( -\frac{\delta_c ^2}{2\sigma (m)^2} \right)
\left(\frac{d \sigma(m)}{dm} \right) dm \label{pss}
\end{equation}
where $\bar{\rho}$ is the present mean density of the universe,
$\sigma(m)$ is the present-day {\em rms} fluctuations of the linear
density field after smoothing with a spherical top-hat filter
containing a mean mass $m$ (and it is specified by the power spectrum,
$P(k)$, of the density fluctuations), and $\delta_c$ is a density
threshold usually taken to be the extrapolated linear overdensity of a
spherical perturbation at the time it collapses. The parameter
$\delta_c$ is only weakly-dependent on 
matter density parameter 
$\Omega\sbr{m}$ and 
the cosmological constant 
$\Omega\sbr{\Lambda}$ (varying by less than 0.02 for flat models with
$\Omega\sbr{m} > 0.1$, see Fig 1. of Eke, Cole, \& Frenk 1996), and so
has been set to its Einsten-de Sitter cosmology value 1.686.

The dimensionless power spectrum ($\Delta(k)\equiv V (2 \pi^2)^{-1}
k^3 |\delta_{k}|^2$ in the linear regime is given by the analytic
approximation of Bond \& Efstathiou (1984)
\begin{equation}
\Delta(k)=\frac{A k^4}{\Big[ 1+
[aq+(bq)^{3/2}+(cq)^2]^{\nu}\Big]^{2/\nu}}
\end{equation}
where $q=k/\Gamma$, $a=6.4 h_{100}^{-1}$ Mpc, $b=3 h_{100}^{-1}$ Mpc, $c=1.7
h_{100}^{-1}$ Mpc, $\nu=1.13$, and $h_{100} = H_0/(100 \kms \mpc^{-1})$.
The parameter $\Gamma=\Omega\sbr{m} h_{100}$ (if we neglect the
effect of baryons) determines the shape, and the normalization is
determined e.g.  by Cosmic Microwave Background fluctuations or by
fixing the {\em rms} fluctuations on a scale of $8 h_{100}^{-1}$ Mpc,
$\sigma_8$.

\subsection{Results}

Figure \ref{fig2} shows the PS-predicted luminosity functions obtained
assuming constant mass--to--light ratio for four different
cosmological models: Standard CDM ($\Omega\sbr{m}=1$,
$\Omega\sbr{\Lambda}=0$, $H_0=50\kms$ Mpc$^{-1}$, $\sigma_8 = 0.51$),
$\Lambda$CDM ($\Omega\sbr{m}=0.3$, $\Omega_{\Lambda} = 0.7$,
$H_0=70\kms$ Mpc$^{-1}$, $\sigma_8=0.9$), $\tau$CDM
($\Omega\sbr{m}=1$, $H_0=50\kms$ Mpc$^{-1}$, $\sigma_8 =0.51$, $\Gamma
=0.21$) and Open CDM ($\Omega\sbr{m} =0.3$, $\Omega\sbr{\Lambda}=0$,
$H_0= 70\kms$ Mpc$^{-1}$, $\sigma_8 = 0.85$).  These models are
described in detail by Jenkins et al.\ (1998) and are normalized
adopting 
the values for the fluctuations in an 8 $h^{-1}_{100}$ Mpc sphere, 
$\sigma_8$, 
prescribed by Eke, Cole, \& Frenk (1996) from their analysis of the
local cluster X-ray temperature function.

The NOG VSLF and the corresponding MFW results are shown in Figure
\ref{fig2}.  The latter is represented by a hatched region, which
shows how the VSLF varies if the overdensity contrast used in
identifying groups is varied from 50 to 300. 

\vbox{%
\begin{center}
\leavevmode
\hbox{%
\epsfxsize=8.9cm
\epsffile{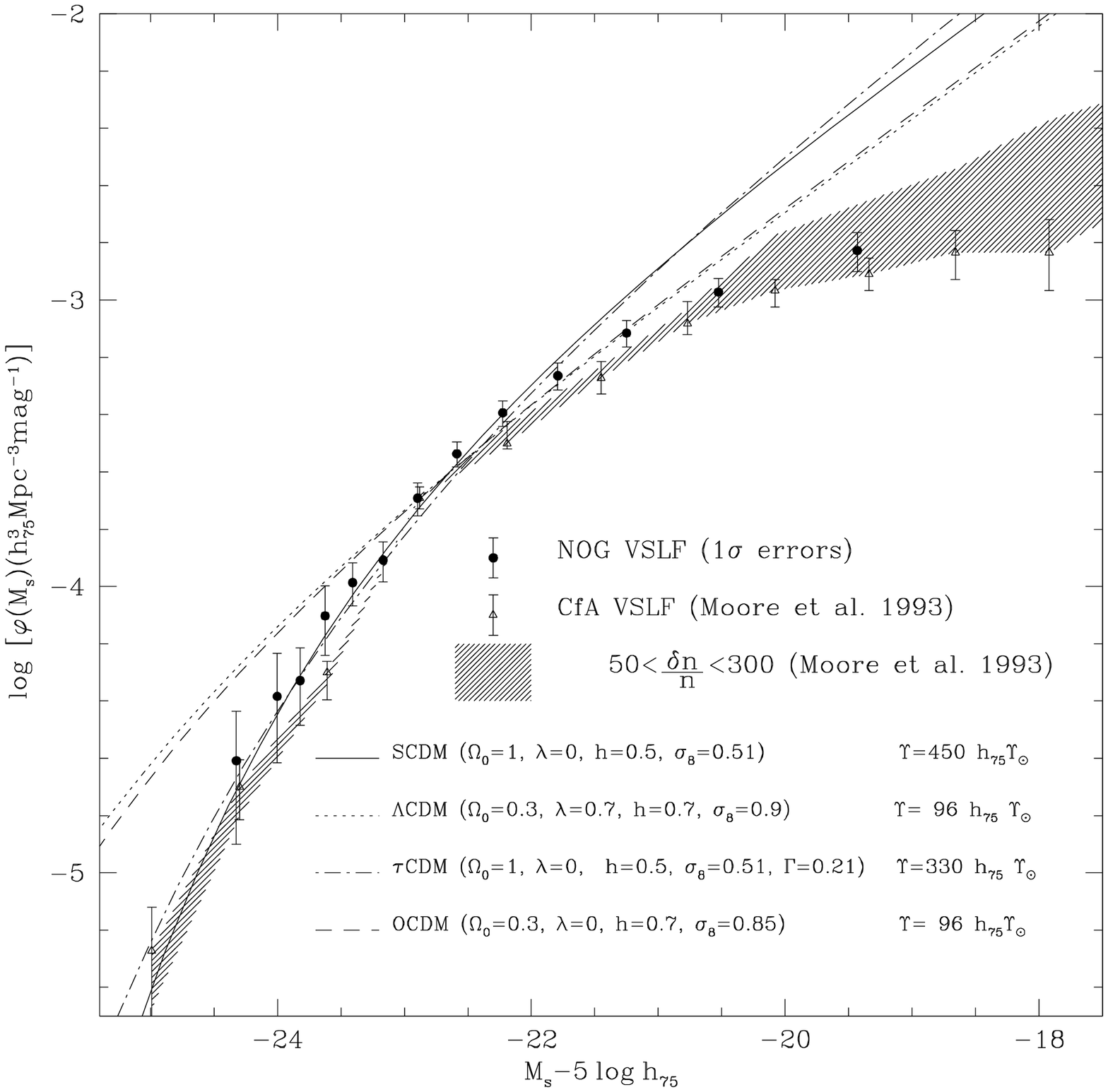}}
\begin{small}
\figcaption{%
The NOG and CfA system LFs are compared with the luminosity
distribution predicted using the Press-Schechter function and a
constant mass--to--light ratio.  Four different cosmological models
(the standard CDM, $\Lambda$CDM $\tau$CDM and open CDM) are computed
with cosmological parameters as given in the figure. For each model we
show the constant mass--to--light ratio adopted.  The hatched region
indicates how the LF varies if the overdensity criterion $\frac{\delta
n}{n}$ used in identifying groups is varied from 50 to 300.
\label{fig2}}
\end{small}
\end{center}}

 The friends-of-friends
grouping algorithm used to define the groups and hence the VSLF
requires that the local density contrast at the edge of a group be
greater than some limiting threshold $\frac{\delta n}{\bar{n}}$ (see
the discussion in \S 2).  If we assume a spherical halo with a
singular isothermal density profile $n(r)\propto r^{-2}$, this local
density threshold corresponds to a mean overdensity
$\langle\frac{\delta n}{\bar{n}}\rangle=3\frac{\delta n}{\bar{n}}$
where the average is performed over the typical separation of group
members (virialization region).  Thus, if galaxy biasing is
negligible, for our grouping algorithms, this implies a value which is
close to the mean overdensity of nearly 180 predicted for a top-hat
spherical collapse in the case of $\Omega\sbr{m}=1$ (see Eke, Cole, \&
Frenk 1996 for values in different cosmologies).

Cole \& Lacey (1996)
found that the virialized regions of N-body halos are reliably
identified using the percolation method and a local density threshold
parameter of nearly 60.  Using this limiting density contrast, they
are able to reproduce the bulk properties of virialized structures (in
particular the virial mass) and concluded that the closeness to global
virial equilibrium of the identified N-body halos depends rather
weakly on the adopted limiting threshold.  Moreover, as noted by
Diaferio et al.\ (1999), the average luminosities of groups
objectively identified from a redshift survey with typical {\em
friends-of-friends} parameters are in agreement with those of groups
extracted from real-space simulations.  Thus, we can use our VSLF for
a meaningful comparison with the PS predictions.

Figure \ref{fig2} indicates that, in agreement with previous results,
all models fail to describe the observed faint-end slope. 

\vbox{%
\begin{center}
\leavevmode
\hbox{%
\epsfxsize=8.9cm
\epsffile{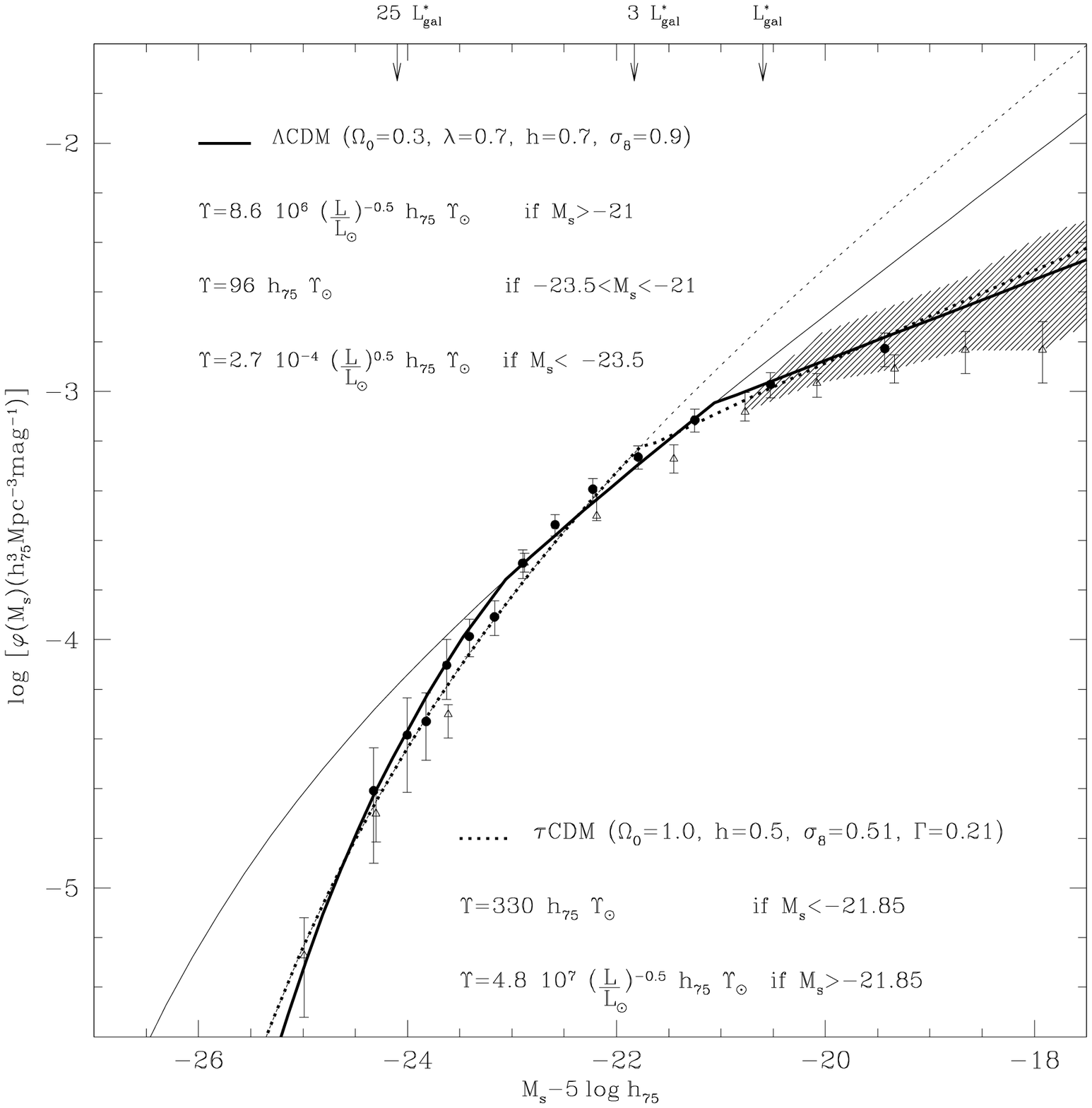}}
\begin{small}
\figcaption{%
The NOG and CfA system LFs are compared with the luminosity
distribution predicted using the Press-Schechter function. The LF in
the $\tau$CDM and $\Lambda$CDM scenarios are computed for constant
mass--to--light ratios ($\Upsilon=330 h_{75}\Upsilon_{\odot}$ and
$\Upsilon=96 h_{75}\Upsilon_{\odot}$ respectively) and also (heavy
lines) using the 
piecewise 
mass--to--light ratios given in the figure.
The hatched region indicates how the LF varies if the overdensity
criterion $\frac{\delta n}{n}$ used in identifying groups is varied
from 50 to 300.
\label{fig3}}
\end{small}
\end{center}}

 More
interestingly, several models ($\Lambda$CDM and OCDM) fail to
reproduce the bright end of the LF, if a constant mass--to--light
ratio is assumed.

We can turn this problem around, and solve for the mass--to--light
function, $\Upsilon(m)$, required to fit the observed VSLF.  We
assume that there is a monotonic relation between the mass and light
of virialized systems, i.e. $\ls = f(m)$ with no scatter.

It is  possible to obtain an exact numerical solution by noting  
that the $\Upsilon$ function
is defined implicitly by the requirement that the number
density of systems above with masses greater than mass $m$ must be the
same as the number density of systems with luminosities greater than
$\ls=f(m)$,
\begin{equation}
N\sbr{s}[>\,\ls=f(m)] = N\sbr{s}(>m)
\label{num}
\end{equation}
where $N\sbr{s}(>\ls)$ is given by the appropriate integral over the VSLF
\begin{equation}
N\sbr{s}(>\ls) = \int_{\ls}^{\infty} \phi(\ls) d\ls
\end{equation}
and $N\sbr{s}(>m)$ is the corresponding integral over the PS mass
function.  

The LF integral can be obtained analytically, but the PS mass function
integral must be solved numerically. The mass--to--light ratio is then
given by
\begin{equation}
\Upsilon = \frac{m}{f(m)}
\end{equation}

\vbox{%
\begin{center}
\leavevmode
\hbox{%
\epsfxsize=8.9cm
\epsffile{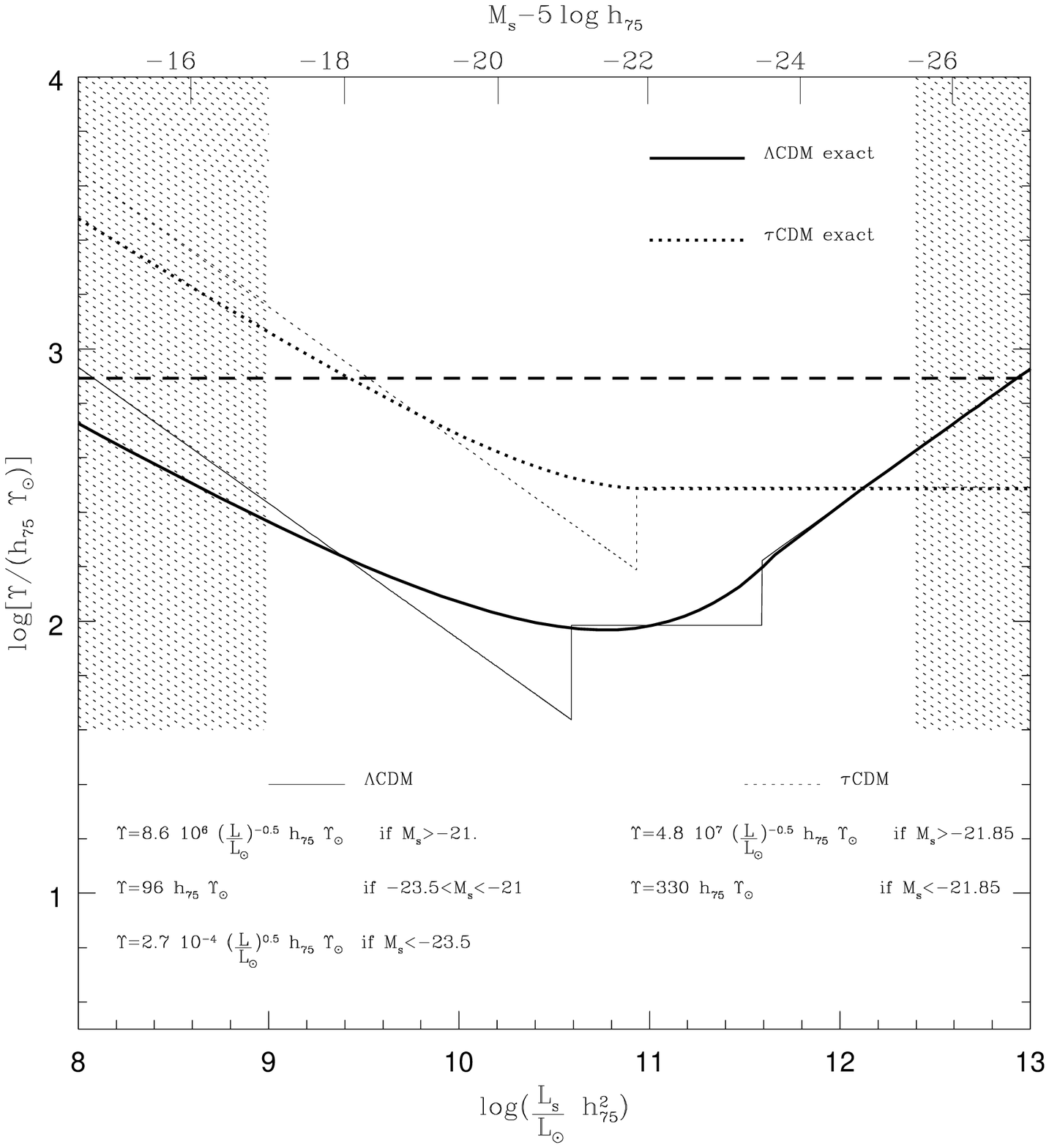}}
\begin{small}
\figcaption{%
Mass--to--light ratio as a function of luminosity for the $\Lambda$CDM
and $\tau$CDM models.  The thin solid and dotted lines indicate the
mass--to--light ratios for the 
piecewise 
models shown in Fig.\ \ref{fig3} for
the $\Lambda$CDM and $\tau$CDM models respectively, whereas the thick
lines are the exact solutions obtained by solving eq.\ \ref{num}.  The
dashed line indicates the critical value of the mass--to--light
ratio. The hatched region represents extrapolation to regions not
covered by our data.
\label{fig4}}
\end{small}
\end{center}}

If we assume that mass and light are simply related via a power law expression  
$\Upsilon \propto L_{s}^{\gamma}$,                                     
the halo luminosity function  assumes the simple analytical form       

\begin{equation}                     
\phi(M_s)\propto m n(m)(1+\gamma)    
\end{equation}                       

The predicted VSLFs obtained fitting           
``piecewise''  $\Upsilon$ functions are shown in Figure \ref{fig3} for the $\tau$CDM
and $\Lambda$CDM models, which are used here as our reference models.

In Table 2, we 
give 
the masses assigned to luminous systems 
for the numerical solution of the $\Upsilon$ function. 
In 
Figure \ref{fig4},
the piecewise and numerical solutions for 
$\Upsilon$ 
are 
plotted as a function of luminosity.
The result is that, for both models, $\Upsilon$ must go approximately
as $\ls^{-0.5 \pm 0.06}$ on small scales.  Interestingly, this
behavior agrees with the direct determinations of $\Upsilon$ derived
for spiral galaxies from the analysis of their rotation curves and for
ellipticals from a variety of tracers of galactic gravitational field
(see the review by Salucci \& Persic 1997).

The $\Upsilon$ function reaches a minimum 
at a  luminosity of approximately 
$10^{10.5}\;h^{-2}_{75}\;L_{\odot}$, which corresponds roughly to
$\lgal^*$.  Then it remains quite flat over $\sim 1$ dex in
luminosity.  The behavior at the bright end is model-dependent: for
the $\tau$CDM model, $\Upsilon$ must remain constant ($m \propto
\ls^{1\pm 0.1}$), whereas for the $\Lambda$CDM model it must increase
with luminosity as $\ls^{0.5 \pm 0.26}$ from the scale of galaxy
groups ($m \sim   10^{13} m_{\odot}\;h_{75}^{-1}$) to that of
rich clusters ( $\sim 10^{15} m_{\odot}\;h_{75}^{-1}$).

\vbox{%
\begin{center}
\leavevmode
\hbox{%
\epsfxsize=8.9cm
\epsffile{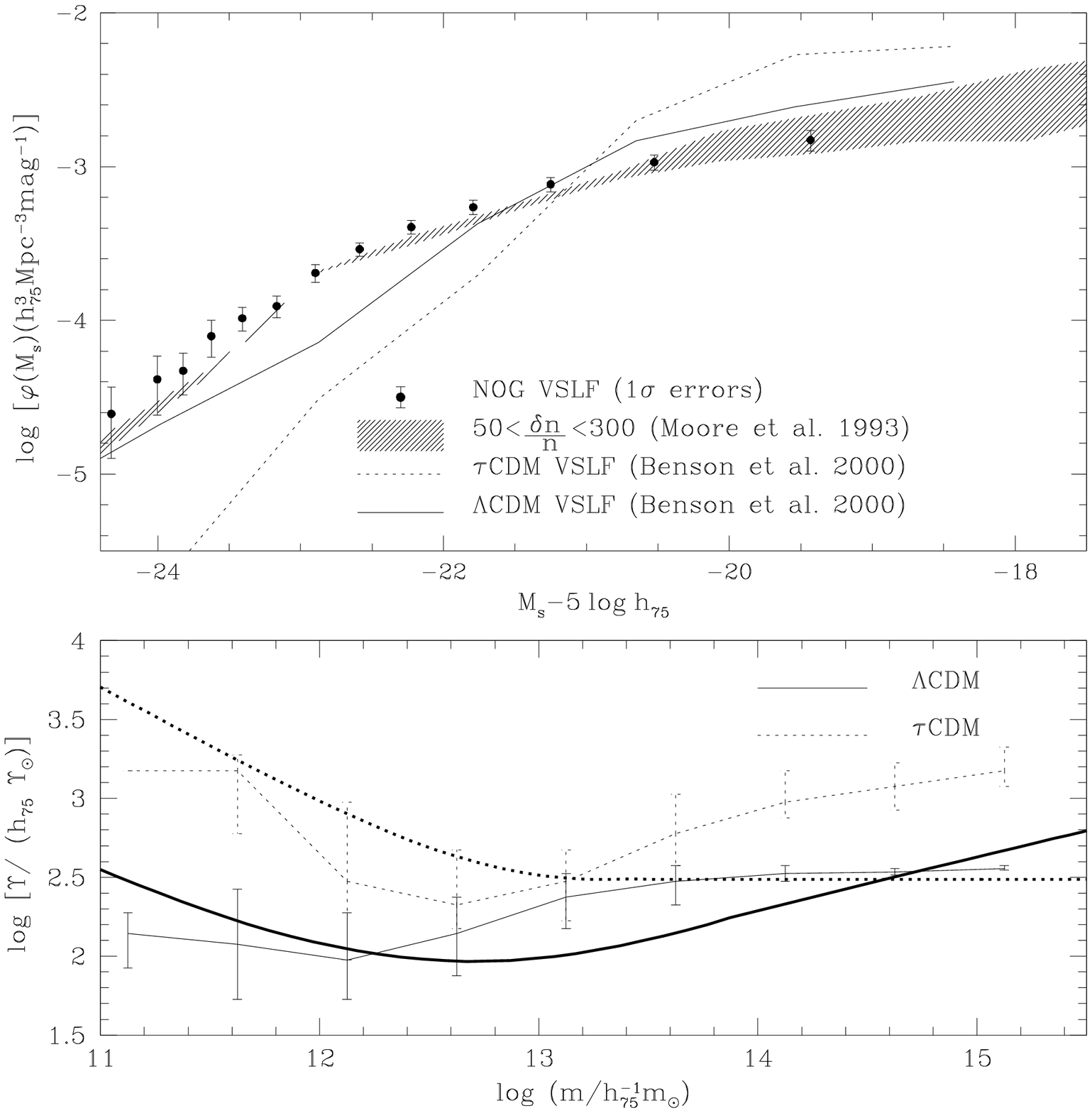}}
\begin{small}
\figcaption{%
{\it Upper:} the NOG VSLF (points with $\pm 1
\sigma$ error bars) is shown in comparison with the semi-analytic
prediction of Benson et al.\  (2000).  Results from their $\tau$CDM and
$\Lambda$CDM models are plotted with short-dashed lines and solid
lines respectively.  {\it Lower:} mass--to--light ratios from Fig. \ref
{fig4} (smooth curves) are compared to those derived by Benson et
al.\ (curves with error bars).
\label{fig5}}
\end{small}
\end{center}}

Recall that our mass--to--light ratios are based on the PS mass
functions. If, as mentioned above, the PS prescription underestimates
the abundance of high-mass objects, the true scaling at the
high-luminosity end is likely to be somewhat steeper than found above.
This would also be true if we used a lower value $\delta_c=1.3 $ in
the PS mass function, as might be expected for a non-spherical
collapse for bound structures.

\subsection{Discussion}

It is interesting to compare the $\Upsilon$ functions derived in the
previous subsection with values found via direct mass estimates. As
discussed above, these direct methods are problematic for low mass
systems such as single galaxies (because the full extent of the dark
matter halo is not probed) or poor groups (because of the small
numbers of dynamical test particles).  The $\Upsilon$ values found in
the literature are consistent with the range $\sim 100 h_{75}
\Upsilon_{\sun}$ to $\sim 300 h_{75} \Upsilon_{\sun}$ spanned by the
$\Lambda$CDM and $\tau$CDM models considered here.

For high masses, we can compare our results with the scaling behavior
of $\Upsilon$ obtained via direct determinations of the virial masses
of galaxy clusters.  Schaeffer et al.\ (1993) found that
$\Upsilon_V\propto L_V^{0.3}$.  Girardi et al.\ (2000) analyzed 105
rich clusters and found $\Upsilon_{B_J}\propto L_{B_J}^{0.2-0.3}$.

They also noted that a variation of $\Upsilon$ with scale could not be
explained by a higher fraction of spirals in poorer clusters, thus
suggesting that a similar result would also be found by using R-band
galaxy magnitudes.  Carlberg et al.\ (1996) found that the number of
bright galaxies per unit mass is systematically lower in cluster with
higher velocity dispersion.  This result would imply that $\Upsilon$
increases with cluster mass, if the shape of the luminosity function
of cluster galaxies were independent of cluster mass.

These results are in agreement with the mass-to-light function of the
$\Lambda$CDM model, but not with that of the $\tau$CDM model.

On the largest scales, the behavior of the mass-to-light function is
less clear.  Virialized objects of supercluster mass are expected to
be extremely rare --- there are certainly none in the NOG volume. It
is possible to obtain direct estimates of the global mass-to-light
ratio can obtained from studies of peculiar velocities, which probe
deviations from the uniform expansion of the Universe.  Note that the
peculiar velocity of a given galaxy is generated by a large number of
massive halos over a range of distances, in both overdense and
underdense regions. Thus the mass-to-light ratio determined by such
studies is an average over a range of mass scales.  Hudson (1993,
1994) studied the density field and peculiar velocity field in a
volume very similar to the NOG.  He assumed a constant mass-to-light
ration for all systems and found that the average mass-to-light ratio
of optical galaxies was 30\% of the mass-to-light ratio of a critical
density Universe, if galaxies are unbiased tracers of the mass. It
would be interesting to see how these conclusions would be altered if
one adopted the mass-to-light function found here for the $\Lambda$CDM
model.

Our results can also be used to determine the efficiency of star
formation, namely the fraction of all baryonic material which has
turned into stars.  We can convert our $\Upsilon$ function into a
stellar-to-total mass ratio if we assume a stellar mass-to-light ratio
averaged over spheroids and disks, of $\Upsilon\sbr{star} = 3.4
\Upsilon_{\sun}$ (Fukugita, Hogan \& Peebles, 1998) in the $B$-band.
Near the minimum of the $\Upsilon$ function, this yields a mass
fraction of stars, $M\sbr{star}/M\sbr{tot} \sim 0.035 h^{-1}_{75}$.
The mass fraction in stars can then be converted into the fraction of
baryons in stars, if we assume that the baryonic-to-total mass
fraction in virialized systems is the same as the global value,
$\Omega\sbr{baryon}/\Omega\sbr{m}$. Using $\Omega\sbr{baryon} = 0.022
h_{100}^{-2}$ from Netterfield et al.\ (2001), we obtain the
stellar-to-baryonic fraction
\begin{equation}
M\sbr{star}/M\sbr{baryon} \sim 0.26 
\left(\frac{\Upsilon}{100 h_{75} \Upsilon_{\sun} }\right)^{-1}
\left(\frac{\Omega\sbr{m}}{0.3}\right)
h_{75} 
\end{equation}
Thus in those dark matter halos which have luminosities in the range
$1 - 10 \lgal^*$, masses in the range $10^{12} - 10^{13} m_{\sun}$ and
where $\Upsilon \sim 100 h_{75} \Upsilon_{\sun}$, $\sim 25$\% of the mass is
converted into stars.  This is considerably more efficient than the
global stellar fraction of $\lesssim 10\%$ (Fukugita et al. 1998).

For the $\Lambda$CDM model, the mass-to-light-ratio increases to
$\Upsilon = 350 h_{75} \Upsilon_{\sun}$ at cluster masses $\sim
5\ten{14} h_{75}^{-1} m_{\sun}$.  At this mass, the star formation
efficiency has dropped to 7.4\%.  If we also allow for the redder
population of such systems, and assign a higher stellar $M/L_B = 4.5$
(Fukugita et al.\ 1998), the star formation efficiency would be
10\%. This is in agreement with the results of Balogh et al.\ (2001).
For the $\tau$CDM model, the mass-to-light ratio is $\sim
330 h_{75} \Upsilon_{\sun}$ for all systems with masses greater than
$10^{13} h_{75}^{-1} m_{\sun}$, 
and in this mass range
the star formation efficiency would be $\sim 25$\%.

\vbox{%
\begin{center}
\leavevmode
\hbox{%
\epsfxsize=8.9cm
\epsffile{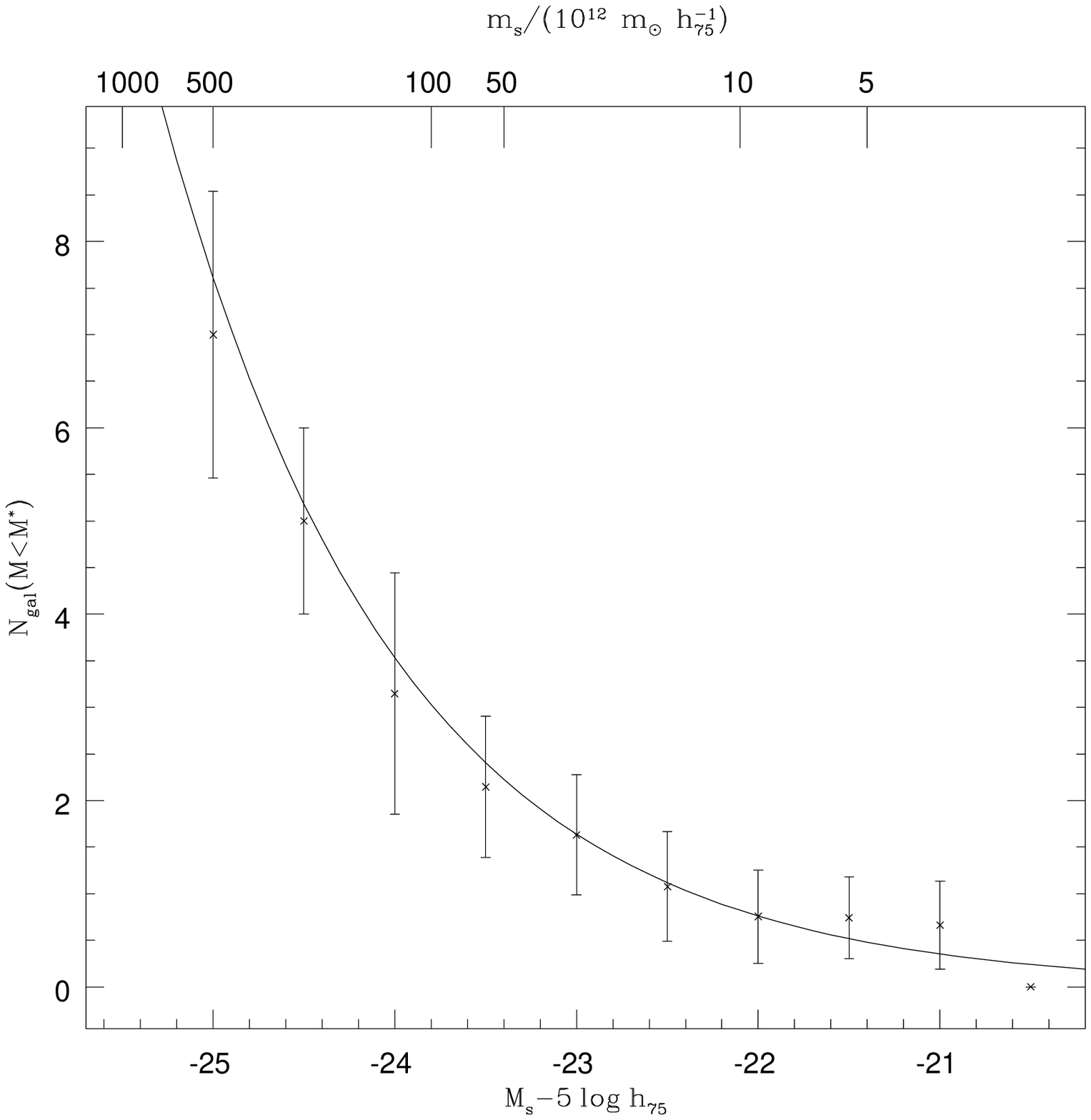}}
\begin{small}
\figcaption{%
The number of galaxies with luminosity greater than $L^{*}\sbr{gal}$
hosted in a halo is plotted as a function of the system luminosity.
Bars represent $\pm 1 \sigma$ errors calculated as the standard
deviation in each bin of absolute magnitude. The mass scale is
obtained using the mass--to--light ratio derived in a $\Lambda$CDM
cosmogony in \S 4. Also shown the best fitting curve as determined in
equation ref{nl}.
\label{fig6}}
\end{small}
\end{center}}

Finally, we can compare our VSLF to the predictions of semi-analytic
models, such as those of Benson et al.\ (2000).  In Figure \ref{fig5},
we show our LF and that of MFW in comparison with the predictions of
their $\Lambda$CDM and $\tau$CDM models.  Benson et al.\ constrained
their models to match the amplitude of the ESP galaxy LF in the
$B_J$-band (Zucca et al.\ 1997) at their $\lgal^*$.  Their $\tau$CDM
model is quite a poor fit, failing to match the slope of our VSLF at
intermediate luminosities and predicting too many low-luminosity
objects.  Their $\Lambda$CDM model provides a marginal fit to our
results: the bright-end slope is correct, although the predicted LF
slope in the range of groups is somewhat too steep.  This arises
because their $\Upsilon$ function increases as $m^{0.3}$ between
masses of $10^{12} m_{\odot}$ and $10^{14} m_{\odot}$, whereas we find
that the behavior of $\Upsilon$ is quite flat in this range (see
Fig. \ref{fig4}). An increase in $\Upsilon$ from the scales of large
galaxies up to clusters is also predicted by the galaxy formation
models of Kauffmann et al.\ (1999), with $\Upsilon_B \propto m^{0.2}$
over the range $10^{12} m_{\odot} < m < 10^{15} m_{\odot}$ for both
their $\Lambda$CDM and $\tau$CDM models. This is similar to that found
by Benson et al., and suggests that the Kauffmann et al.\ VSLF would
be qualitatively similar to that of Benson et al.  The $\Upsilon$
functions obtained from the semi-analytic models of Somerville et
al. (2001) also increase with increasing mass, however their
$\Upsilon$ values are larger than those of Benson et al. and are thus
in poorer agreement with those derived here.

\section{Halo Occupation Numbers}

For many purposes, it is important to know the {\em number\/} of
galaxies which populate a given halo. Once this function is known, one
can use the well-understood clustering of halos to obtain the
clustering of galaxies (Neyman, Scott and Shane 1953; Kauffmann et
al.\ 1999; Diaferio et al.\ 1999; Benson et al.\ 2000; Seljak 2000;
Peacock and Smith 2000; Scoccimarro et al.\ 2001; Benson 2001).

In Fig. \ref{fig6}, we plot the number of galaxies with luminosity
greater than $\lgal^{*}$ as a function of the corrected halo
luminosity and of the total halo mass derived under the assumption of
a $\Lambda$CDM cosmogony (\S 3). Note that NOG sample is complete for
$\lgal > \lgal^*$ (see Fig. 6 of Paper V) and no luminosity selection
effects are polluting the observed trend.  Note that, by definition,
we are only considering the high-luminosity ($\lgal>\lgal^*$) end of
the $\Upsilon$ function.  We find that the number of galaxies above
the $\lgal^*$ luminosity threshold, scales with the total luminosity
of the system as follows
\begin{equation}
N\sbr{gal}(>\lgal^*)=(0.26 \pm 0.02) \left(\frac{\ls}{\lgal^*}\right)^{
0.83 \pm 0.25}\,, \label{nl}
\end{equation}
for $M\sbr{s} - 5 \log h_{75} < -21$.  Using the mass--to--light ratio
derived in section \S 3 we can derive the dependence of the number of
objects residing within halos in terms of the halo mass and of the
specific cosmological model adopted.  We find that in a $\Lambda$CDM
and $\tau$CDM cosmogonies, the scaling of the halo occupation number
with mass is given by the following expressions
\begin{equation}
N\sbr{gal}(>\lgal^*)=6.3 h_{75}^{4/3} \cdot 10^{-8}
\left(\frac{m}{m_{\odot}} \right)^{0.55 \pm 0.26}  \,
\end{equation}
for $m \gtrsim 5\ten{12} h^{-1}_{75} m_{\odot}$ and 
\begin{equation}
N\sb{gal}(>\lgal^*)=4.2 h_{75} \cdot 10^{-12}
\left(\frac{m}{m_{\odot}} \right)^{0.83 \pm 0.33}
\end{equation}
for $ m \gtrsim 10^{13} h^{-1}_{75} m_{\odot}$, respectively.

Peacock and Smith (2000) performed a similar analysis using CfA
(Ramella, Pisani \& Geller 1997) and ESO Slice Project groups (Ramella
et al. 1999).  Their Figure 6 suggests power-law exponents of $\sim
0.55$ and $\sim 0.7$, for $\Lambda$CDM and $\tau$CDM models
respectively, in good agreement with our results. 

\section{The relation between optical and X-ray properties in
virialized systems}

Having derived the functional form of the optical luminosity
distribution of virialized systems, we now investigate the relation
between optical and X-ray emission properties in galaxy systems. The
techniques involved are the same as those developed in section \S 2.
The general problem of deriving the functional form $\lx=f(L)$,
relating the optical and the X-ray luminosities, can be formulated
using the following differential equation
\begin{equation}
\left\{
\begin{array}{l}
{\cal L}\sbr{X}=f({\cal L}\sbr{s}) \\ \\
\frac{df(\ls)}{d\ls}=\frac{\phi(\ls)}{\phi(\lx)}
\end{array}
\right.
\label{de}
\end{equation}
where $\phi(\ls)$ is the optical LF of systems emitting in the X-ray
band with a luminosity distribution $\phi\sbr{X}(\lx)$, and where
${\cal L}\sbr{X}=f({\cal L}\sbr{s})$ is the 
initial condition,
which can be determined from the data.

\vbox{%
\begin{center}
\leavevmode
\hbox{%
\epsfxsize=8.9cm
\epsffile{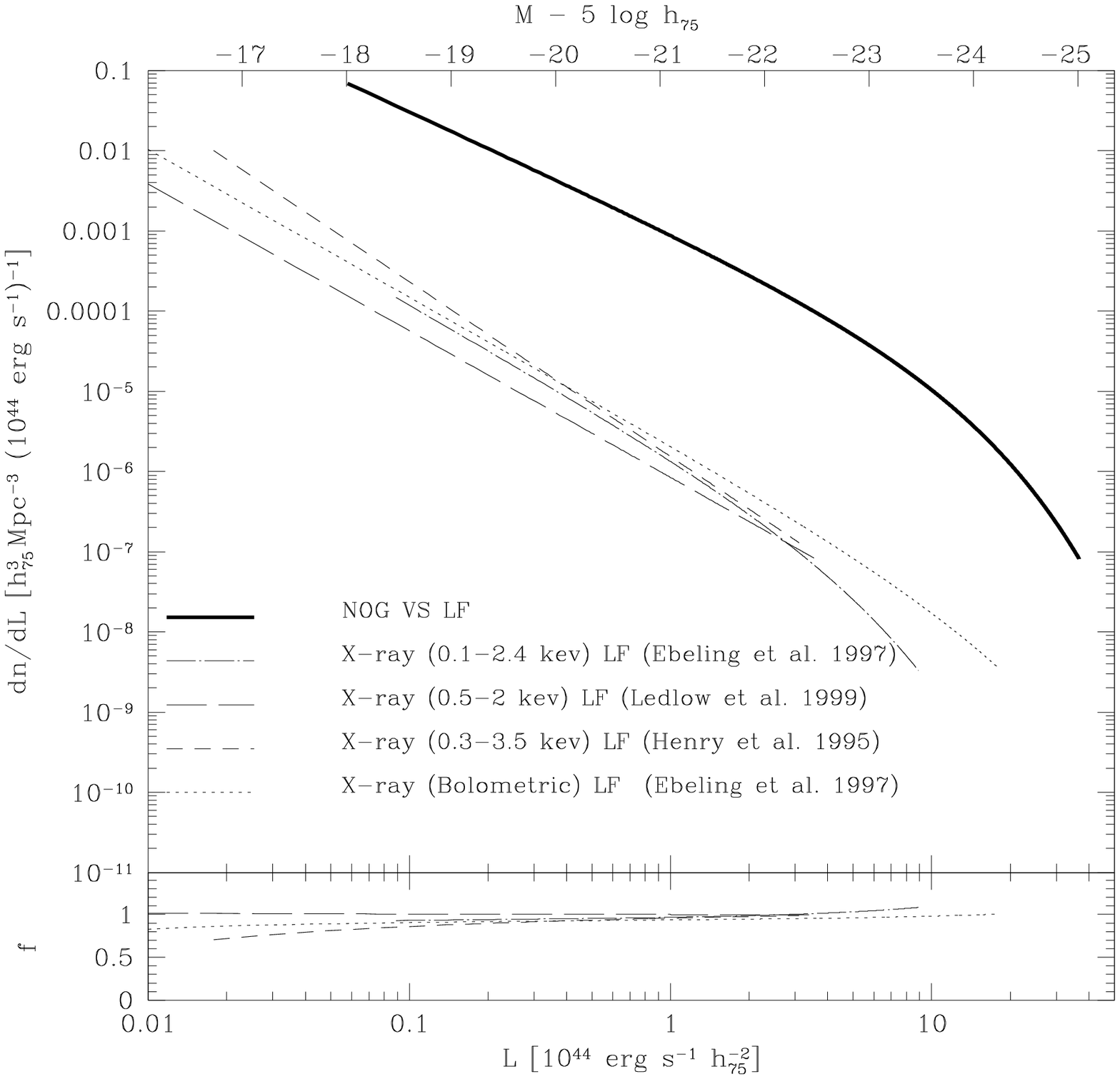}}
\begin{small}
\figcaption{%
{\it Upper:} the local X-ray luminosity
functions in four different bands are compared over a luminosity range
that describes both groups (faint end) and clusters (bright end). The
shape of the NOG VSLF is shown for comparison.  {\it Lower:} the
ratio f between the logarithms of the four X-ray luminosity functions
and a power law expression of parameters $\alpha\sbr{X}=-1.85$ and
$\phi^{*}\sbr{X} = 4 \cdot 10^{-7} Mpc^{-3} (10^{44} {\rm erg s}^{-1}
)^{\alpha\sbr{X}-1}$ is shown as a function of luminosity.
\label{fig7}}
\end{small}
\end{center}}

We assume that all virialized systems detected in the optical also
radiate in X-rays. Their optical LF is a Schechter function whose
parameters $(\phi\sbr{s}^{*},\alpha\sbr{s}, M\sbr{s}^{*})$ are known
over the luminosity range $ -18 < M\sbr{s}-5 \log h_{75} < -25$ (\S 3
and Fig.  \ref{fig7}). The XLFs in various bands are given by Ebeling
et al.\ (1997).

The optical emission is expected to be connected with the X-ray
emission under the standard assumption that the same gravitational
potential shapes the gas density distribution as well as the galaxy
distribution (Cavaliere \& Fusco-Femiano 1976).  We can gain some
insights into the optical-to-X-ray luminosity ratio making some
approximations. Fig. \ref{fig7} shows that $\phi\sbr{X}(\lx)$, for
both groups and clusters, can be well approximated, in almost all the
X-ray bands, by a single power-law expression
$\phi\sbr{X}(\lx)=\phi^{*}\sbr{X}\;L^{\alpha\sbr{X}}$.  In particular,
in the bolometric and 0.1-2.4 keV bands, the local XLFs of Ebeling et
al.\ (1997) are very well approximated (and extrapolated into the
group domain) by the following fitting parameters $\phi^{*}\sbr{X}
\sim 4\cdot 10^{-7}$ Mpc$^{-3} (10^{44}\;{\rm erg}\;{\rm s}^{-1}
)^{\alpha\sbr{X}-1}$ and $\alpha\sbr{X} \sim 1.85$. In this case, the
particular solution of eq. \ref{de} is simple and can be expressed by
the following analytical formula
\begin{equation}
\lx=\Big\{ f({\cal L}\sbr{s})^{\beta\sbr{X}}-C \beta\sbr{X} \Big[ \Gamma_i
\Big(\beta\sbr{s},\frac{\ls}{L_{*}}\Big) -\Gamma_i \Big(\beta\sbr{s},\frac{{\cal
L}\sbr{s}}{L_{*}} \Big) \Big] \Big\}^{\frac{1}{\beta\sbr{X}}} \label{fa}
\end{equation}
where $C=\phi\sbr{s}^{*}/\phi\sbr{X}^{*}$, $\beta=1+\alpha$ (for s
and X subscripts) and $\Gamma_i$ is the incomplete gamma function.
However, if we integrate numerically the differential equation
\ref{de}, we obtain the result plotted in Fig. \ref{fig8}.

\vbox{%
\begin{center}
\leavevmode
\hbox{%
\epsfxsize=8.9cm
\epsffile{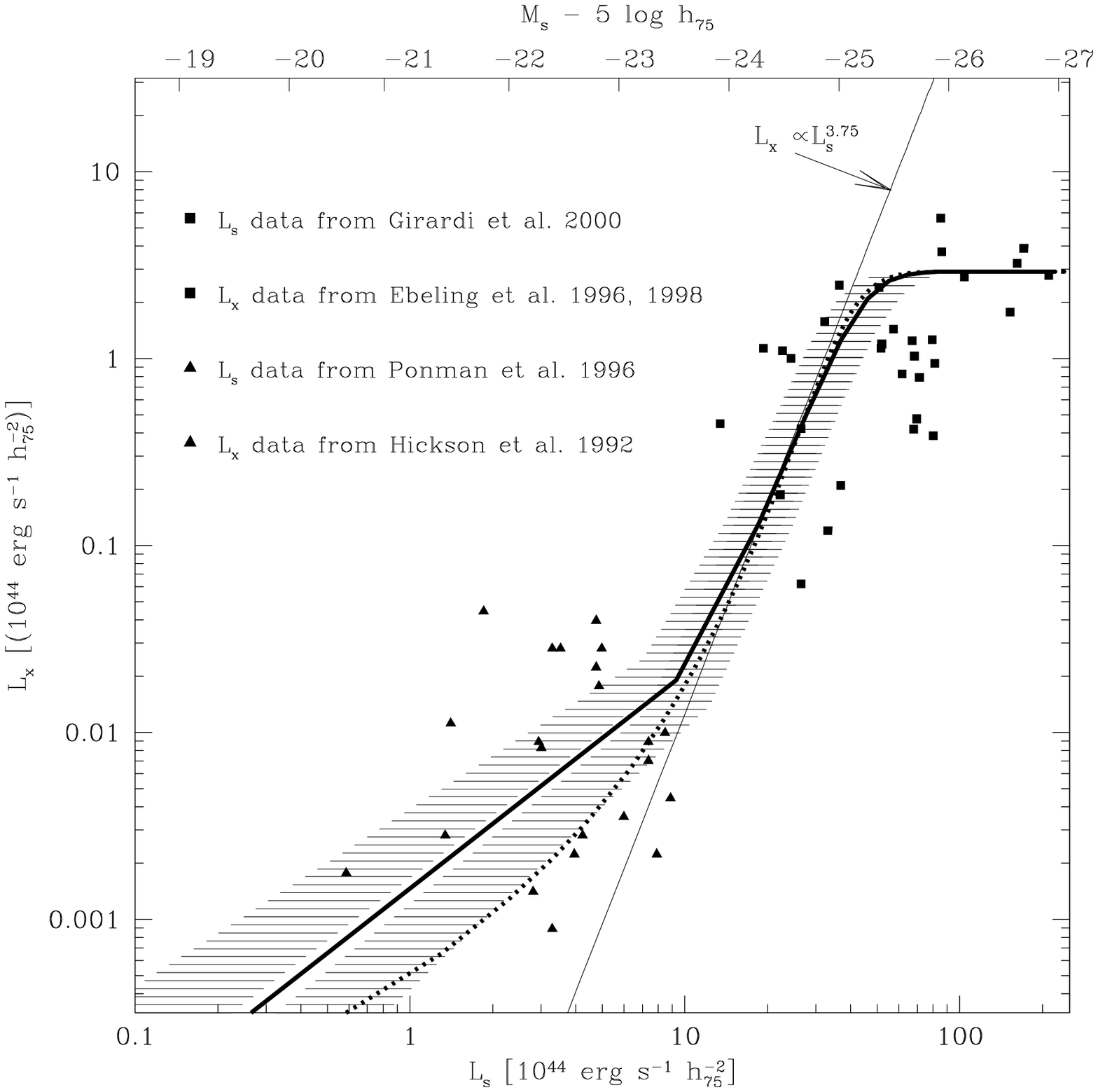}}
\begin{small}
\figcaption{%
The relation between optical ($\ls$) and X-ray
($\lx$) luminosities as determined in eq. \ref{fa} is shown together
with the observed luminosities of Abell clusters. The straight line is
the scaling relation predicted in a $\Lambda$CDM universe using the
optical (this paper) and the X-ray (Ledlow et al.\ 1999)
mass--to--light ratios with arbitrary normalization.  The thick solid
line is the exact solution and the dotted line is the analytical
solution of eq. \ref{fa}.  The hatched region corresponds to the
errors in the $\lx-\ls$ relation obtained from the errors in the
respective luminosity functions.
\label{fig8}}
\end{small}
\end{center}}

In order to test the solution and determine the normalization
constant, we have used a sample of Abell clusters for which both the
optical blue magnitudes and the X-ray luminosities in the 0.1-2.4 keV
band are known.  The largest local samples of X-ray luminosities
compiled to date are the X-ray Brightest Abell Cluster Sample XBACS
(Ebeling et al.\ 1996) and the Brightest Cluster Sample (BCS) (Ebeling
et al.\ 1998).  We found that 28 XBACS and 3 BCS clusters also have
estimated $B_J$ luminosities (Girardi et al.\ 2000).  At the
low-luminosity end we used a sample of compact groups whose
luminosities in the X-ray (bolometric) and optical (blue) passbands
are given by Ponman et al.\ (1996) and Hickson et al.\ (1992)
respectively.  In order to reduce observational errors, the initial
condition has been determined by averaging the luminosities of the
four optically brightest clusters in the sample.

The overall agreement between our prediction and the data is shown in
Fig.  \ref{fig8}.  Note that the Girardi et al.\ (2000) data shown
here are at the tail of our optical VSLF. Of particular interest is
the apparent break in the power-law behavior at $M\sbr{s} = -21 + 5
\log h_{75}, L\sbr{X} \sim 10^{42} {\rm erg\ s}^{-1} h_{75}^{-2}$.
Fainter than this break, the numerical solution is well approximated by
$\lx\propto \ls^{1.5}$, whereas in the brighter regime, both numerical
and analytic solutions are well approximated by $\lx\propto
\ls^{3.5}$.

\section{Scaling relations in groups and clusters of galaxies}

When considering X-ray properties, as well as mass-to-light ratios, it
is convenient to separate poor systems, by which we mean systems with
one or a few $\lgal^*$ galaxies (poor groups), from rich systems (rich
groups and clusters).  We define the former to be those systems with
$-21 > M\sbr{s} - 5\, \log \, h_{75} > -23$ or $\lx < 10^{42}\;{\rm erg\
s}^{-1} h_{75}^{-2}$, and the latter to be systems with $-23 >
M\sbr{s} - 5\, \log \, h_{75} > -25$ or $ 10^{42} < \lx / ({\rm erg\
s}^{-1} h_{75}^{-2}) < 5 \cdot 10^{44}$.  Beyond this range the X-ray
and optical LFs are not well determined from our data

\subsection{Scaling relations for rich systems}

In the regime of rich systems, as defined above, the X-ray-to-optical
scaling is $\lx \propto \ls^{3.5}$.  In this regime, for a
$\Lambda$CDM cosmology, we found in \S 4 that
\begin{equation}
m \propto \ls^{1.5 \pm 0.26 } \label{mlr}\,.
\end{equation}
This then implies that $m \propto \lx^{0.43\pm0.08}$. This is in good
agreement with the analysis of Ledlow et al.\ (1999), who compared the
observed XLF to PS mass functions and found $m \propto L\sbr{X}^{0.4
\pm 0.03}$, in the bolometric X-ray band over the range $10^{41} < \lx/
({\rm erg\ s}^{-1} h_{75}^{-2})<5\cdot 10^{45}$) (solution
$\Lambda$CDM2 in their Table 1 which is similar to the $\Lambda$CDM
model adopted here.)

It is possible to compare this scaling with other cluster parameters
such as velocity dispersion $\sigma$ and temperature $T$.  Under
conditions of spherically-symmetric and isothermal equilibrium, the
X-ray luminosity is connected to the velocity dispersion of the
virialized halos by the relation $\lx\propto f^2 \sigma^3 T^{1/2}$
(Quintana \& Melnick 1982), where $f$ is the ratio of the gas mass to
the total cluster mass.  If the gas and galaxies are in hydrostatic
equilibrium, the average plasma temperature is proportional to the
depth of the cluster potential well ($T \propto \sigma^2$), leading to
the following simple scaling relation $\lx\propto f^2 \sigma^4$.  The
observed $\lx - \sigma$ relation for clusters is not very far from
this theoretical prediction. Several authors reported an empirical
$\lx - \sigma$ relation close to $\lx \propto \sigma^4$ (Quintana \&
Melnick 1982; Mulchaey and Zabludoff 1998), while Xue \& Wu (2000) and
White, Jones \& Forman (1997) derived somewhat steeper relations, $\lx
\propto \sigma^{5.3\pm0.21}$ and $\lx \propto \sigma^{6.38 \pm 0.46}$,
respectively.

Our results allow an alternative determination of this scaling
relationship.  Since our systems are defined by a fixed overdensity
criterion, their typical size scales as $R\sbr{s} \propto
[\int_{L\sbr{min}}^{\infty} L \Phi(L) dL]^{-1/3}$.  Using eq. 5 of
paper V and the cosmology dependent relation $m=L^{\alpha}$ we obtain,
\begin{equation}
\ls \propto \sigma^{\frac{6}{3\alpha -1}} \,.
\label{lssig}
\end{equation}
Using the observed scaling relationship for the X-ray--to--optical
luminosity ratio ($\lx \propto \ls^{\beta}$) we can write
\begin{equation}
\lx \propto \sigma^{\frac{6\beta}{3\alpha -1}}
\end{equation}
For rich systems in the $\Lambda$CDM cosmology, this yields
$\ls\propto \sigma^{1.7 \pm 0.38}$ which is in good agreement with the
results of Schaeffer et al.\ (1993), who found $L_V\propto
\sigma^{1.87\pm 0.44}$ and those of Adami et al.\ 1998 who reported
$L_{B\sbr{J}} \propto \sigma^{1.56}$ with a large scatter.

In the X-ray band, our result would translate into the following
scaling relation, $\lx=\sigma^{6 \pm 1.3}$, consistent with those
found by Xue \& Wu (2000) and White, Jones, \& Forman (1997).  In
$\tau$CDM cosmology we would have obtained the following relations
$\ls\propto \sigma^{3 \pm 0.13}$ and $\lx=\sigma^{10.5 \pm 1.3}$, in
disagreement with the observations.

We can go a step further if we assume that clusters follow an
isothermal relation $\sigma \propto T^{1/2}$ (which appears to be
supported by the data of Ponman et al.\ 1996; White, Jones, \& Forman
1997; Xue \& Wu 2000).  Inserting this $\sigma-T$ model into the
$\lx-\sigma$ scaling relationship we obtain $\lx \propto T^{3 \pm
0.65}$. Again this result is in agreement with the fit of White,
Jones, \& Forman (1997) and Xue \& Wu (2000).

Thus, in the regime of rich systems, the derived scalings are in
agreement with the observations for the $\Lambda$CDM cosmology.  For
the $\tau$CDM cosmology the agreement is poorer.

\subsection{Scaling relations for poor systems}

For poor systems, i.e. super-$\lgal^*$ galaxies and poor groups, we
found $\lx \propto \ls^{1.5}$. In this regime the mass--to--light
ratio is nearly constant, $m \propto \ls^{1\pm 0.1}$, for both
$\Lambda$CDM and $\tau$CDM cosmologies.  This yields $\lx \propto
\ls^{2.2\pm0.3}$, which, besides being in agreement with what we
found, confirms in an independent way the bivariate scaling behavior
of the optical--to--X-ray luminosity ratio over the group and cluster
scales.  Extending this analysis to velocity dispersions using
\ref{lssig}, we derive a steep relationship $\ls \propto \sigma^{3 \pm
0.45}$.  Now coupling our results with the observationally determined
$\lx-\ls$ scaling relationship we obtain $\lx \propto \sigma^{3.9 \pm
1.2}$.

Because of the problems mentioned in \S 1, direct observational
results for groups and poor clusters are much scattered, although they
tend to lead to a flatter $\lx - \sigma$ relation compared to that of
clusters.  Mulchaey \& Zabludoff (1998) obtain $\lx\propto \sigma^{4.3
\pm 0.4}$, Ponman et al.\ (1996) obtain $\lx\propto \sigma^{4.9
\pm2.1}$, Helsdon \& Ponman (2000) give $\lx \propto \sigma^{4.5 \pm
1.1}$, whilst Mahdavi et al.\ (1997) and Mahdavi et al.\ (2000) found
much flatter relations, i.e.  $\lx \propto \sigma^{1.56 \pm 0.25}$ and
$\lx \propto \sigma^{0.4 \pm 0.3}$.  The fit of Xue \& Wu (2000) is
intermediate in slope with $\lx \propto \sigma^{2.35\pm0.21}$.  Our
derived relation is consistent with the steep slopes and is marginally
consistent with that of Xue \& Wu.

Now if we assume $\sigma \propto T$ (Helsdon \& Ponman 2000), we
obtain the steeper relation $\lx \propto T^{3.9 \pm 1.2}$ which is in
agreement with the theoretical predictions of Cavaliere, Menci \&
Tozzi (1997, 1999) (see the discussion at the end of the section) and
in excellent agreement with the results $\lx \propto T^{4.9 \pm 0.9}$
of Helsdon \& Ponman (2000).

If we had assumed a condition of virial equilibrium between condensed
and diffuse baryons on the scale of groups, we would have obtained a
flatter relation $\lx \propto T^{2.5}$ similar to the trend observed
in clusters.  Since this does not provide a good fit to data (see
Fig. \ref{fig9}), we reinforce the conclusion of Ponman et al.\ that
at the low end hierarchy of the clustering pattern, i.e. for low X-ray
temperature systems, the virial equilibrium condition between galaxies
and gas does not apply.

\subsection{Discussion}

The break in power-law scalings of mass, $\lx$, $\sigma$ and $T$
between rich and poor systems, occurring at $\lx \sim 10^{42} {\rm erg
s}^{-1} h_{75}^{-1}$, has been noted by many authors. Here we have
found that a change in the shape of the $\lx-\ls$ relation occurs at
the same point. 

\vbox{%
\begin{center}
\leavevmode
\hbox{%
\epsfxsize=8.9cm
\epsffile{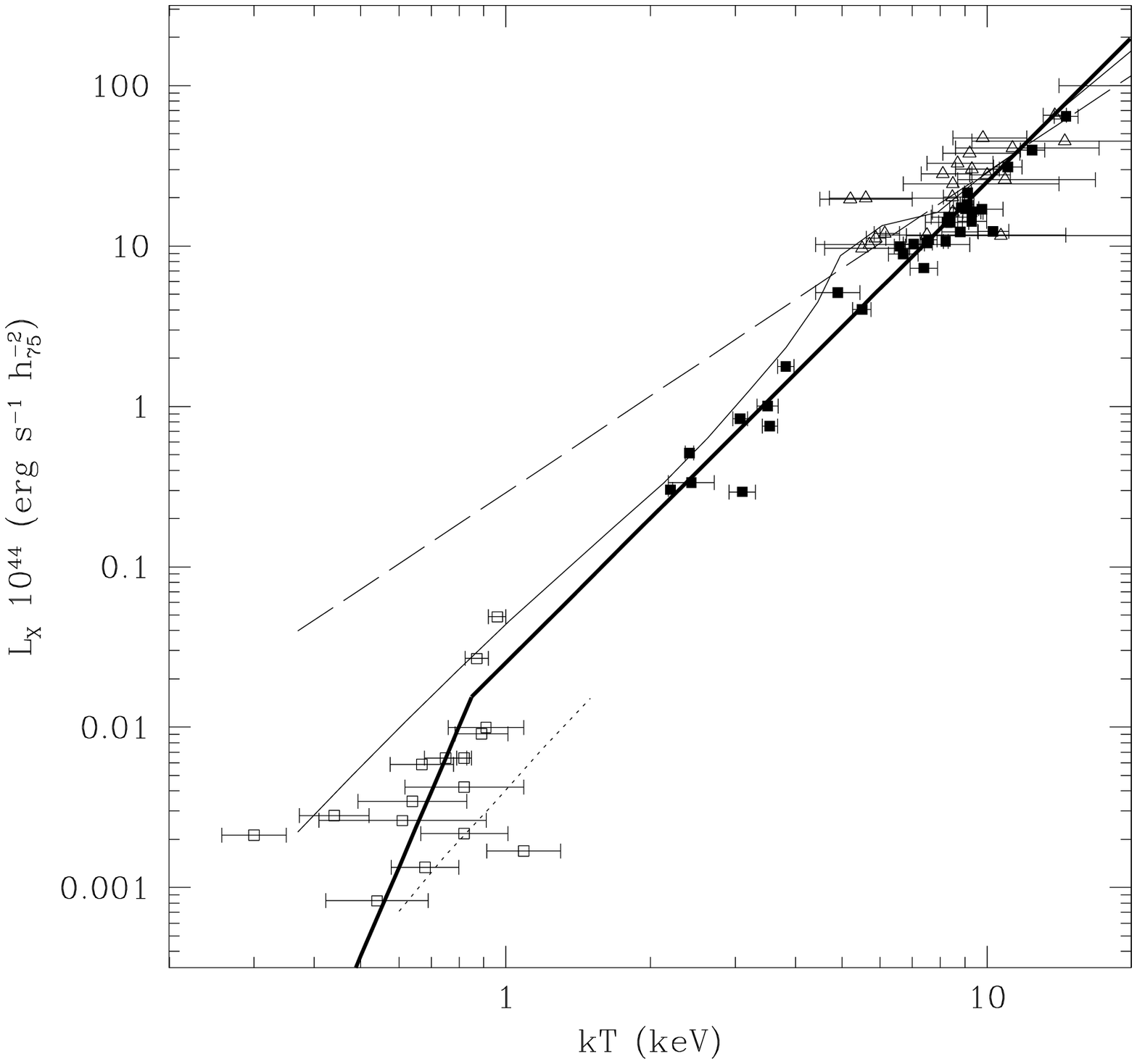}}
\begin{small}
\figcaption{%
The relation between the bolometric luminosity and the temperature in
$\Lambda$CDM cosmology. Data are from Arnaud \& Evrard (1999) (filled
squares), Allen \& Fabian (1998) (triangles) and Ponman et al.\ (1996)
(empty squares). The thick line refers to our predicted scaling
behavior.  The dashed line refers to the self-similar case and the
solid line to predictions of the thermodynamic model of the $z=0$ ICM
(in a $\Lambda$CDM scenario) developed by Tozzi \& Norman (1999) (the
entropy excess is $K=0.3 \cdot 10^{34} {\rm erg} {\rm cm}^2 {\rm
g}^{-5/3}$ constant with epoch and the cooling is included).  The
dotted line shows the relation $\lx-T$ at z=0 derived by Tozzi and
Norman (1999) within the projected radius used in Ponman et al.\
(1996).
\label{fig9}}
\end{small}
\end{center}}

 At a similar point there is a net change in the
chemical properties and the spatial distribution of the ICM on the
scales of groups (Renzini 1997, 1999).  One might think that this is
due solely to a change in the properties of the ICM.  However, if the
$\Lambda$CDM PS mass function is correct, then it is interesting that
this transition corresponds to the same mass scale ($M\sbr{s} \sim -23
+ 5 \log h_{75}$) which separates efficient galaxy formation, with
$\Upsilon$ low and approximately constant, from inefficient galaxy
formation at which $\Upsilon$ begins to rise as $\ls^{0.5}$.  This
would suggest a connection between galaxy formation and the properties
of the ICM.  There are two proposed models for this connection.  The
star formation may have preheated the gas, or the low-entropy gas may
have preferentially cooled into stars.

The broken power-law behavior is in quantitative agreement with
predictions of the preheating scenario (Kaiser 1991, Evrard \& Henry
1991) in which the entropy of the hot, diffuse intracluster medium is
raised at early time (prior to gravitational collapse) by
non-gravitational heating such as feedback effects of star formation,
SN winds, shocks etc.  Cavaliere, Menci, \& Tozzi (1997, 1999)
proposed a semi-analytic model of shocked ICM gas to explain the the
observed $\lx-T$ relation.  The same relation is reproduced by a model
of Balogh et al.\ (1999) using the physics of an adiabatic isentropic
collapse of a preheated gas.  Both models make detailed predictions
for the $\lx-T$ relations which are approximated by the power laws
$\lx \propto T^{5}$ in the luminosity range $\lx < 0.25 \cdot 10^{44}
({\rm erg\ s}^{-1} h_{75}^{-2})$ (groups) and $\lx \propto T^{3}$ in
the luminosity range $0.25\cdot 10^{44} \leq \lx / ({\rm erg\ s}^{-1}
h_{75}^{-2}) \leq 5 \cdot 10^{44}$ (rich groups and clusters).  Tozzi
\& Norman (2000) combine these two scenarios in a single model taking
into account an initial entropy excess and the transition between the
adiabatic and shock regime in the growth of X-ray halos.  Their
$\lx-T$ model and our predicted scalings are compared to data in
Fig. \ref{fig9}.

On the other hand, the preferential cooling of low-entropy gas into
stars in galaxies might also break the scaling (Thomas \& Couchman
1992; Bryan 2000; Muanwong et al.\ 2001).  For this process to explain
the scaling, one requires greater efficiency of star formation for
low-mass halos, e.g. 10\% at $T \sim 1$ keV vs 4\% at $T \sim 10$ keV
(Bryan 2000). As discussed in \S 4, for the $\Lambda$CDM models, we
found that the maximum star formation efficiency was as high as 25\%
for poor systems, about 2.5 times the global efficiency of 10\%.  For
rich systems the star formation efficiency drops below the global
average. Note that for a different mass spectrum, e.g. $\tau$CDM there
is no mass dependence of the star formation efficiency for high
masses.

In summary, at least for the $\Lambda$CDM spectrum, it is interesting
that the break in X-ray scaling properties occurs at roughly the same
mass as the change in star formation efficiency.  This suggests that
there is a physical link between galaxy formation and the X-ray
properties of the hot gas.  Of course increased galaxy formation will
lead to greater feedback as well as more efficient cooling of
low-entropy gas, so it is difficult to distinguish between the
scenarios discussed above.

\section{Summary and Conclusions}

We have used the B-band luminosity function of virialized systems to
investigate scaling relations over a wide dynamic range of mass, from
single galaxies to clusters of galaxies.

When our LF is compared to the Press-Schechter mass functions
predicted in CDM cosmogonies we find that all these models fail, if a
constant mass--to--light ratio is assumed.  Specifically, in the
$\tau$CDM and $\Lambda$CDM models, the mass--to--light ratio must vary
as $L^{-0.5}$ to match the faint-end of the luminosity function.  On
the other hand, a constant mass--to--light ratio and a ratio varying
as $L^{0.5}$ match the bright ends of the $\tau$CDM and $\Lambda$CDM
models, respectively.  For the $\Lambda$CDM model, the efficiency of
star formation is maximized in the range of one to $\sim 5$ $\lgal^*$
or $10^{12.5}$ to $10^{13.5} h^{-1}_{75}$ $m_{\sun}$, the regime of
single galaxies and poor groups. In this range, the fraction of
baryons is stars is $\sim 25\%$. The latter behavior of the
mass--to--light ratio is in qualitative agreement with the predictions
of recent semi-analytical models of galaxy formation, in which galaxy
formation is inhibited by the reheating of cool gas on small-mass
scales and by the long cooling times of hot gas on large-mass scales.
The variation of the mass--to--light ratio with scale found by some
authors through direct estimates of the masses of galaxy systems lends
indirect support to the $\Lambda$CDM cosmology. A further test of the
varying mass--to--light ratio model could be made via the comparison
of observed peculiar velocities with the predictions from nearly
all-sky optical galaxy catalogs such as the NOG.

Since our sample of galaxies is complete for objects brighter than
$\lgal^*$, we have also measured the halo occupation number.  We find
that, for $\Lambda$CDM models, this quantity grows more slowly than
the total mass of the system, as required in the ``halo'' model of
explain clustering.

Comparing X-ray luminosities as a function of optical luminosities in
virialized systems, we find a break in the power-law slope of the
relation as we go from groups to clusters, independent of the
cosmological model.  Dynamical scaling relations in X-ray systems are
known to have break at a similar mass scale. Furthermore, the break
occurs at a similar mass scale to that at which the mass-to-light ratio
changes in the $\Lambda$CDM model.  
This suggests physical link between galaxy formation and the
X-ray properties.  However, these data cannot say whether this is due
to reheating of the hot halo gas by supernovae or efficient cooling of
low-entropy gas.

We look forward to forthcoming large field redshift surveys planned
with SDSS (Gunn \& Knapp 1993), 2dF (Colless 1998), 2MASS (Huchra et
al.\ 1998), which will (a) extend the range over which the system
luminosity is determined (b) lower the luminosity threshold over which
group members are selected and (c) reduce the errors in their
distribution statistics. Moreover the DEEP2 redshift survey (Davis \&
Faber 1998) will probe the variation with cosmic time of the $M/L \, -
\,L$, $\lx \, - \, L$ and $L \, - \, \sigma$ relations, allowing a
deeper understanding of galaxy formation.

\acknowledgements

We wish to thank C.\ M.\ Baugh, A.\ J.\ Benson, M. Davis, R.\
Giovanelli, M.\ Haynes, P.\ Monaco, J.\ Newman, E.\ Scannapieco, P.\ Tozzi, for
interesting conversations.  We are especially indebted to P. Tozzi who
provided us results in advance of publication.

CM acknowledges the NSF grant AST-0071048,  MJH
acknowledges a grant from the NSERC of Canada.

\clearpage

\begin{deluxetable}{cccc}
\tablewidth{0pc}
\tablecaption{Masses of  systems for different models \label{tab1}}
\tablehead{
\colhead{$ M\sbr{s}-5 \log h_{75}$}  &
\colhead{$ \frac{\ls}{L_{\odot}}h_{75}^2$} &
\colhead{$ \frac{m\sbr{s}}{m_{\odot}}h_{75}$} &
\colhead{$ \frac{m\sbr{s}}{m_{\odot}}h_{75}$}  \\
\colhead{ } &
\colhead{ }  &
\colhead{($\tau$CDM)} &
\colhead{($\Lambda$CDM)} }
\startdata
-18&$2.5 \times 10^9   $&$2.0  \times 10^{12}$&$ 4.2 \times 10^{11}$ \\
-19&$6.2 \times 10^9   $&$3.5  \times 10^{12}$&$ 8.2 \times 10^{11}$ \\
-20&$1.6 \times 10^{10}$&$6.6  \times 10^{12}$&$ 1.7 \times 10^{12}$ \\
-21&$3.9 \times 10^{10}$&$1.3  \times 10^{13}$&$ 3.7 \times 10^{12}$ \\
-22&$9.8 \times 10^{10}$&$3.0  \times 10^{13}$&$ 9.4 \times 10^{12}$ \\
-23&$2.5 \times 10^{11}$&$7.6  \times 10^{13}$&$ 3.0 \times 10^{13}$ \\
-24&$6.2 \times 10^{11}$&$1.9  \times 10^{14}$&$ 1.2 \times 10^{14}$ \\
-25&$1.6 \times 10^{12}$&$4.8  \times 10^{14}$&$ 5.1 \times 10^{14}$ \\
-26&$3.9 \times 10^{12}$&$2.1  \times 10^{15}$&$ 2.1 \times 10^{15}$ \\
\enddata
\end{deluxetable}
\clearpage

\end{document}